\documentclass[twocolumn,showpacs,preprintnumbers,amsmath,amssymb,nofootinbib]{revtex4-1}

\usepackage{graphicx}
\usepackage{dcolumn}
\usepackage{bm}
\usepackage{amsfonts}
\usepackage{color}

\def\bequ{\begin{equation}}
\def\eequ{\end{equation}}
\def\be{\begin{equation}}
\def\ee{\end{equation}}

\begin{document}
\title{Charged Dirac perturbations on Reissner-Nordstr\"om black holes in a cavity: \\quasinormal modes with Robin boundary conditions
}

\author{Jia Liu}
\author{Mengjie Wang}
\email{Corresponding author: mjwang@hunnu.edu.cn.}
\author{Zishuo Wang}
\author{Haoyu Liu}
\author{Jinshan An}
\author{Jiliang Jing}
\affiliation{\vspace{2mm}{
Department of Physics, Key Laboratory of Low Dimensional Quantum Structures and Quantum Control of Ministry of Education, Hunan Research Center of the Basic Discipline for Quantum Effects and Quantum Technologies, and Institute of Interdisciplinary Studies, Hunan Normal University, Changsha, Hunan 410081, P.R. China\vspace{1mm}}}
\date{\today}

\begin{abstract}
We investigate charged Dirac quasinormal spectra on Reissner-Nordstr\"om black holes in a mirror-like cavity. For this purpose, we first derive charged Dirac equations, and \textit{two} sets of Robin boundary conditions following the vanishing energy flux principle. The Dirac spectra are then computed both analytically and numerically. Our results reveal a symmetry hidden in the Dirac spectra between two boundary conditions. Moreover, when the cavity is placed close to the event horizon $r_+$, we identify that, in the neutral background the Dirac spectra asymptote to $-(3/8+N/2)i$ [$-(1/8+N/2)i$] for the first [second] boundary condition; while in the charged background the real part of charged Dirac spectra asymptote to $qQ/r_+$ for both boundary conditions; where $N$ is the overtone number, $q$ and $Q$ are charges for the field and for the background. In particular, we uncover a striking anomalous decay pattern, $i.e.$ the excited modes decay \textit{slower} than the fundamental mode, when the charge coupling $qQ$ is large. Our results further illustrate the robustness of vanishing energy flux principle, which are applicable not only to anti-de Sitter black holes but also to black holes in a cavity.
\end{abstract}
\maketitle

\section{Introduction}
Black holes (BHs) are among the most mysterious compact objects in the universe, created by the collapse of matter into an object with strong gravitational force~\cite{Misner:1973prb}. The existence of BHs has been directly confirmed by BH images and gravitational waves, see for example~\cite{EventHorizonTelescope:2019dse,EventHorizonTelescope:2019ths,EventHorizonTelescope:2019ggy,EventHorizonTelescope:2022wkp,Chen:2022scf,Chen:2023wna,LIGOScientific:2016aoc,LIGOScientific:2018mvr,LIGOScientific:2016sjg,LIGOScientific:2017bnn,KAGRA:2021vkt,Jing:2023vzq,Jing:2023vzq,Jing:2022vks,Jing:2021ahx,Zou:2021lkj} and references therein. Since BHs always interact with their environments, the stability of BHs becomes one of central issues in BH physics. When a BH is perturbed, oscillations of the BH are characterized by a superposition of damped exponentials with complex frequencies, which are the well known quasinormal modes (QNMs). The quasinormal spectra contain information of the classic and even quantum properties of a BH spacetime~\cite{Davis:1971gg,Hod:1998vk,Andersson:2003fh,Cardoso:2008bp,Cardoso:2016rao,Konoplya:2018yrp,Oshita:2023cjz} and, therefore have been attracted much attention, see excellent reviews~\cite{Kokkotas:1999bd,Nollert:1999ji,Berti:2009kk,Konoplya:2011qq,Berti:2025hly} and references therein.

Technically QNMs are defined as the eigenvalue of the perturbation equation, with specific physically motivated \textit{boundary conditions}. For BHs in an asymptotically flat spacetime, which is the most relevant case in the gravitational wave era~\cite{Barack:2018yly}, the boundary conditions are normally imposed as an ingoing wave boundary condition at the event horizon and an outgoing wave boundary condition at infinity~\cite{Chandrasekhar:1975zza}. While for a system composed of BHs and a mirror-like cavity, the above mentioned boundary conditions cannot be applied anymore.

The BH-mirror system is indeed a physically interesting model, first introduced in~\cite{Press:1972zz} to construct BH bomb. Since then, such system has been widely explored in the context of BH dynamics and especially BH QNMs~\cite{Cardoso:2004nk,Hod:2013fvl,Hod:2014pza,Hod:2016kpm,Dias:2018zjg,Herdeiro:2013pia,Degollado:2013bha,Sanchis-Gual:2015lje,Sanchis-Gual:2016tcm,Sanchis-Gual:2016ros,Dias:2021acy}. To look for QNMs for BHs in a cavity, one normally imposes the Dirichlet boundary condition at the location of the cavity, not only for scalar fields~\cite{Cardoso:2004nk,Hod:2014pza,Hod:2013fvl,Hod:2016kpm} but also for high spin fields~\cite{Hod:2013xde}. Since the Dirichlet boundary condition is \textit{not} valid for high spin fields in asymptotic anti-de Sitter (AdS) BHs and, by considering the similarity between the BH-mirror system and the AdS BH, one may naturally wonder if the Dirichlet boundary condition is still available for high spin fields on BHs in a cavity. This question has been partially answered for Maxwell fields on Schwarzschild BHs in a cavity~\cite{Lei:2021kqv} where the full spectra can only be attained by employing the vanishing energy flux boundary conditions~\cite{Wang:2015goa,Wang:2015fgp,Wang:2016dek,Wang:2017fie,Wang:2019qja,Wang:2021upj,Wang:2021uix}, instead of Dirichlet boundary condition\footnote{More precisely, the Dirichlet boundary condition is also applicable, but only to the Regge-Wheeler variable. Yet only one set of the spectra may be obtained in this situation.}. While the same quasinormal spectra of Maxwell fields may be also obtained by treating the mirror as a conductor~\cite{Brito:2015oca}, indicating the vanishing energy flux boundary conditions are not essential to study Maxwell QNMs.

In order to further verify the robustness of the vanishing energy flux principle for BHs in a cavity as well as to clarify the similarity between AdS BHs and BHs in a cavity, here we apply the vanishing energy flux principle both to neutral and to charged Dirac fields on Reissner-Nordstr\"om (RN) BHs in a cavity. Our results demonstrate the generality of the vanishing energy flux principle and, in particular, reveal a striking anomalous decay pattern of charged Dirac quasinormal spectra. In addition, we identify that QNMs of the BH-mirror system are more sensitive to the BH event horizon, by comparing to the counterparts of AdS BHs.

The structure of this paper is organized as follows. In Section~\ref{seceq} we briefly introduce the RN geometry and derive, for massless charged Dirac fields, the equations of motion and \textit{two} sets of explicit Robin type boundary conditions under the vanishing energy flux principle. In Section~\ref{secana} charged Dirac equations are solved \textit{analytically}, in the low frequency and small charge coupling approximations and when the cavity is placed far away from the BH, in terms of a standard matching method. Beyond the approximations valid for analytic calculations, the matrix method is then employed and various numeric results for both boundary conditions are presented in Section~\ref{secnum}. Final remarks and conclusions are presented in the last section. 

\section{background geometry, field equations and boundary conditions}
\label{seceq}
In this section, we first briefly review the geometry of the four-dimensional RN BH and then, derive both equations of motion for massless charged Dirac fields on this background and the corresponding boundary conditions satisfied by Dirac fields at the location of the cavity.
\subsection{Reissner-Nordstr\"om BHs}
The line element of a four-dimensional RN BH may be written as
\begin{equation}
ds^2=\dfrac{\Delta}{r^2}dt^2-\dfrac{r^2}{\Delta}dr^2-r^2d\theta^2-r^2\sin^2\theta d\varphi^2 \,,\label{metric}
\end{equation}
with the metric function
\begin{eqnarray}
\Delta \equiv r^2-2Mr+Q^2\,,\nonumber
\end{eqnarray}
and the electromagnetic potential of the RN BH is
\begin{equation}
A=-\dfrac{Q}{r}dt\,.\label{potential}
\end{equation}
Here $M$ and $Q$ are physical mass and electric charge of the background. The event horizon $r_+$ and the Cauchy horizon $r_-$ are given by
\begin{equation}
r_{\pm}=M\pm\sqrt{M^2-Q^2}\,,\nonumber
\end{equation} 
where the background charge satisfies $Q\leq M$, and the equal sign may only be achieved for extremal BHs.

\subsection{Charged Dirac equations}
A massless charged Dirac field obeys the equation
\begin{equation}
\gamma^\mu(\mathcal{D}_\mu-\Gamma_\mu)\Psi=0\,,\label{ChargedDiraceqU1}
\end{equation}
where
\begin{equation}
\mathcal{D}_\mu\equiv \partial_\mu-iqA_\mu\,,\;\;\;\Gamma_\mu=-\frac{1}{8}(\gamma^a\gamma^b-\gamma^b\gamma^a)\Sigma_{ab\mu}\,.\label{spincon}
\end{equation}
Here $q$ is the field charge, $A_\mu$ is the electromagnetic potential given by Eq.~\eqref{potential}, $\gamma^a(a=0, 1, 2, 3)$ are the ordinary Dirac matrices defined in the Bjorken-Drell representation~\cite{Bjorken:100769}, and
\begin{equation}
\Sigma_{ab\mu}=e_a^\nu(\partial_\mu e_{b\nu}-\Gamma^\alpha_{\nu\mu}e_{b\alpha})\,,\nonumber
\end{equation}
where the tetrad $e_a^\mu$ may constructed in terms of $\gamma^\mu=e_a^\mu\gamma^a$. The matrices $\gamma^\mu$, associated to the geometry given in Eq.~\eqref{metric}, are defined as~\cite{Wang:2017fie}
\begin{align}
&\gamma^t=\dfrac{r}{\sqrt{\Delta}}\gamma^0\,,\;\;\;\;\;\;\;\gamma^r=\dfrac{\sqrt{\Delta}}{r}\gamma^3\,,\nonumber\\
&\gamma^\theta=\dfrac{1}{r}\gamma^1\,,\;\;\;\;\;\;\;\;\;\;\;\gamma^\varphi=\dfrac{1}{r\sin\theta}\gamma^2\,.\label{gammams}
\end{align}

By taking the following ansatz for Dirac fields
\begin{equation}
\Psi=\left(
\begin{matrix}
\eta\\
\eta
\end{matrix}
\right)\,,\;\;\;\eta=\frac{e^{-i\omega t}e^{im\varphi}}{(r^2\Delta\sin^2\theta)^{1/4}}
\left(
\begin{matrix}
R_1(r)S_1(\theta)\\
R_2(r)S_2(\theta)
\end{matrix}
\right)\,,\label{Chargedfielddecom}
\end{equation}
Eq.~\eqref{ChargedDiraceqU1} leads to a set of first order differential equations which are coupled 
\begin{align}
&\sqrt{\Delta}\left(\dfrac{d}{dr}-\dfrac{iK}{\Delta}\right)R_1(r)=\lambda R_2(r)\,,\label{Chargedfirstorderr1}\\
&\sqrt{\Delta}\left(\dfrac{d}{dr}+\dfrac{iK}{\Delta}\right)R_2(r)=\lambda R_1(r)\,,\label{Chargedfirstorderr2}\\
&\left(\dfrac{d}{d\theta}-\dfrac{m}{\sin\theta}\right)S_1(\theta)=\lambda S_2(\theta)\,,\label{Chargedfirstorderang1}\\
&\left(\dfrac{d}{d\theta}+\dfrac{m}{\sin\theta}\right)S_2(\theta)=-\lambda S_1(\theta)\,,\label{Chargedfirstorderang2}
\end{align}
with $K \equiv \omega r^2-qQr$, where $\omega$ and $m$ are the frequency and azimuthal number of Dirac fields, respectively.

The above first order equations may be decoupled and, through straightforward calculations, one obtains the following second order radial equations 
\begin{align}
&\sqrt{\Delta}\dfrac{d}{dr}\left(\sqrt{\Delta}\dfrac{d}{dr}\right)R_{1}(r)+H_1(r)R_{1}(r)=0\,,\label{ChargedDiracU2radialR1}\\
&\sqrt{\Delta}\dfrac{d}{dr}\left(\sqrt{\Delta}\dfrac{d}{dr}\right)R_{2}(r)+H_2(r)R_{2}(r)=0\,,\label{ChargedDiracU2radialR2}
\end{align}
with
\begin{align}
&H_1(r)=\dfrac{K^2+\tfrac{i}{2}K\Delta^\prime}{\Delta}-iK^\prime-\lambda^2\,,\nonumber\\
&H_2(r)=\dfrac{K^2-\tfrac{i}{2}K\Delta^\prime}{\Delta}+iK^\prime-\lambda^2\,,\nonumber
\end{align}
where $\prime$ and $\lambda$ denote a derivative with respect to $r$ and the separation constant, and $\lambda=1, 2, 3, \cdots$~\cite{Dolan:2009kj}.

The second order radial equations, given in Eq.~\eqref{ChargedDiracU2radialR1}--Eq.~\eqref{ChargedDiracU2radialR2}, will be the main focus of interest in the remaining part of this paper.

\subsection{boundary conditions}
In order to look for Dirac quasinormal frequencies on RN BHs in a cavity, one has to solve the radial equations \eqref{ChargedDiracU2radialR1}--\eqref{ChargedDiracU2radialR2} with \textit{physically motivated} boundary conditions. At the event horizon, we employ the usually applied ingoing wave boundary condition; while at the location of the cavity, we follow the generic principle proposed in~\cite{Wang:2015goa,Wang:2015fgp,Wang:2016dek,Wang:2017fie,Wang:2019qja,Wang:2021upj,Wang:2021uix} by requiring that the \textit{energy flux} of Dirac fields vanishes.

The energy flux through a 2-sphere at radial coordinate $r$ is
\begin{equation}
\mathcal{F}|_r=\int_{S^2} \sin\theta d\theta d\varphi\; r^2 T^r_{\;\;t}\,, \label{Chargedflux1}
\end{equation}
where $T^r_{\;\;t}$ may be obtained straightforwardly from the energy-momentum tensor of charged Dirac fields
\begin{equation}
T_{\mu \nu}=\dfrac{i}{8\pi}\bar{\Psi}\left[\gamma_\mu(\mathcal{D}_\nu-\Gamma_\nu)+\gamma_\nu(\mathcal{D}_\mu-\Gamma_\mu)\right]\Psi+c.c. \,.\label{EMTensorDirac}
\end{equation}
Here $\bar{\Psi}\equiv \Psi^\dag\gamma^0$, $\Psi^\dag$ is the hermitian conjugate of $\Psi$ and $c.c.$ stands for complex conjugate of the preceding terms. 

By substituting Eqs.~\eqref{spincon}--\eqref{Chargedfielddecom} into Eq.~\eqref{EMTensorDirac} and integrating $T^r_{\;\;t}$ over a $2$-sphere with a fixed $r$, Eq.~\eqref{Chargedflux1} becomes\footnote{Alternatively, the same conditions may be derived by calculating the number current $\int_{S^2} \sin\theta d\theta d\varphi\; r^2J^r$, where $J^r=\bar{\Psi}\gamma^r\Psi$.}
\begin{equation}
\mathcal{F}|_r\propto |R_1|^2-|R_2|^2\,,\label{Chargedflux2}
\end{equation}
where the angular functions $S_{1,2}(\theta)$ have been normalized
\begin{equation}
\int_0^\pi d\theta\; |S_{1,2}(\theta)|^2=1\,.\nonumber
\end{equation}

The explicit boundary conditions satisfied by $R_1$, according to Eqs.~\eqref{Chargedfirstorderr1}--\eqref{Chargedfirstorderr2}, may be derived from Eq.~\eqref{Chargedflux2}, which are\footnote{The relative phase between two moduli appeared in Eq.~\eqref{Chargedflux2} has been fixed by calculating normal modes in a pure cavity.} 
\begin{align}
\dfrac{dR_1(r)}{dr}|_{r_m}=i\left(\dfrac{K(r_m)}{\Delta(r_m)}+\dfrac{\lambda}{\sqrt{\Delta(r_m)}}\right)R_1(r_m)\,,\label{R1bc1}\\
\dfrac{dR_1(r)}{dr}|_{r_m}=i\left(\dfrac{K(r_m)}{\Delta(r_m)}-\dfrac{\lambda}{\sqrt{\Delta(r_m)}}\right)R_1(r_m)\,,\label{R1bc2}
\end{align}
where $r_m$ is the location of the cavity, $\Delta(r_m)=r_m^2-2Mr_m+Q^2$, $K(r_m)=\omega r_m^2-qQr_m$. These boundary conditions are obviously \textit{Robin} type and, the existence of two different sets of boundary conditions indicates the existence of two distinct branches of modes. As we will show in the following analytic and numeric calculations that this is \textit{indeed} the case. 

Following exactly the same procedures as above, one may derive the explicit boundary conditions satisfied by $R_2$, which are 
\begin{align}
\dfrac{dR_2(r)}{dr}|_{r_m}=-i\left(\dfrac{K(r_m)}{\Delta(r_m)}+\dfrac{\lambda}{\sqrt{\Delta(r_m)}}\right)R_2(r_m)\,,\label{R2bc1}\\
\dfrac{dR_2(r)}{dr}|_{r_m}=-i\left(\dfrac{K(r_m)}{\Delta(r_m)}-\dfrac{\lambda}{\sqrt{\Delta(r_m)}}\right)R_2(r_m)\,.\label{R2bc2}
\end{align}
By comparing Eqs.~\eqref{R2bc1}--\eqref{R2bc2} to Eqs.~\eqref{R1bc1}--\eqref{R1bc2}, one immediately realizes that they are complex conjugate with each other, which is also held for AdS BHs.

One may look for the Dirac spectra either by solving Eq.~\eqref{ChargedDiracU2radialR1} with boundary conditions~\eqref{R1bc1}--\eqref{R1bc2} for $R_1$ or by solving Eq.~\eqref{ChargedDiracU2radialR2} with boundary conditions~\eqref{R2bc1}--\eqref{R2bc2} for $R_2$. While, as we have checked, the same Dirac spectra are obtained. Therefore, for concreteness and without loss of generality, in the following we focus on $R_1$, $i.e.$ Eq.~\eqref{ChargedDiracU2radialR1} with the corresponding boundary conditions given in Eqs.~\eqref{R1bc1}--\eqref{R1bc2}.

\section{Analytics}
\label{secana}
In this section we calculate \textit{analytically}, by applying the standard \textit{asymptotic matching method} to Dirac fields, both normal modes in a pure cavity and the imaginary part of QNMs on RN BHs in a cavity. The goal of this part is to show \textit{explicitly}, on one hand, that the vanishing energy flux boundary conditions \textit{do} generate \textit{two} different sets of modes, even in a pure cavity; while on the other hand, contradictory to bosonic fields, that charged Dirac fields \textit{cannot} trigger black hole bomb on RN BHs in a cavity. Moreover, such analytic calculations provide initial values for numeric calculations.  

The logic of the asymptotic matching method is as follows. The exterior region of BHs may be divided into the \textit{near} ($r-r_+\ll1/\omega$) and \textit{far} ($r-r_+\gg r_+$) regions, and Eq.~\eqref{ChargedDiracU2radialR1} may be solved in each region when the charge coupling satisfies $qQ\ll1$. With the low frequency approximation ($\omega r_+\ll1$), these two regions are overlapped and the solutions obtained in these two regions are valid in the \textit{overlap} region. Then QNMs can be calculated by imposing physically motivated boundary conditions. By requiring that the cavity is placed far away from the BH ($r_+\ll r_m$), it is then possible to compute quasinormal frequencies perturbatively, on top of normal modes in a pure cavity. 

\subsection{Dirac normal modes in a pure cavity}
In a pure cavity the Dirac equation, given by Eq.~\eqref{ChargedDiracU2radialR1}, is simplified into
\begin{align}
&r\dfrac{d}{dr}\left(r\dfrac{d}{dr}\right)R_1(r)+\left(\omega^2r^2-i\omega r-\lambda^2\right)R_1(r)=0\,,\nonumber
\end{align}
which can be solved as
\begin{align}
R_1(r)=\dfrac{1}{\sqrt{r}}\left(c_1M_{-\frac{1}{2}, \lambda}(2i\omega r)+ c_{2}W_{-\frac{1}{2}, \lambda}(2i\omega r)\right)\,,\label{solpurecavity}
\end{align}
where $c_1$ and $c_2$ are two integration constants and, $M_{\kappa, \lambda}(z)$ and $W_{\kappa, \lambda}(z)$ are Whittaker functions. Expanding Eq.~\eqref{solpurecavity} at the origin and imposing the regularity condition, one shall fix $c_2=0$ so that Eq.~\eqref{solpurecavity} becomes 
\begin{align}
R_1(r)\sim\dfrac{1}{\sqrt{r}}M_{-\frac{1}{2}, \lambda}(2i\omega r)\,.\label{solpurecavity2}
\end{align}

By noticing that, in a pure cavity, BH effects may be neglected so that $\Delta(r_m)=r_m^2$ and $K(r_m)=\omega r_m^2$. Then by substituting Eq.~\eqref{solpurecavity2} into the boundary conditions given in Eqs.~\eqref{R1bc1} and~\eqref{R1bc2}, and by utilizing the property 
\begin{equation}
\frac{d M_{\kappa,\lambda}(z)}{d z}=\big (\frac{1}{2}-\frac{\kappa}{z}\big)M_{\kappa,\lambda}(z)+\frac{1}{z}\big(\kappa+\lambda+\frac{1}{2}\big)M_{\kappa+1,\lambda}(z)\;,\nonumber
\end{equation} 
two sets of modes characterized by $\omega_1$ and $\omega_2$ may be derived, satisfying
\begin{align}
\dfrac{M_{\frac{1}{2}, \lambda}(2i\omega_1 r_m)}{M_{-\frac{1}{2}, \lambda}(2i\omega_1 r_m)}=i\,,\;\;\;\;\;\dfrac{M_{\frac{1}{2}, \lambda}(2i\omega_2 r_m)}{M_{-\frac{1}{2}, \lambda}(2i\omega_2 r_m)}=-i\,.\label{normalmodes}
\end{align}

From Eq.~\eqref{normalmodes}, together with the fixed separation constant $\lambda$ and overtone number $N$, Dirac normal frequencies may be calculated numerically. We list a few selected fundamental normal frequencies for Dirac fields by varying $\lambda$ in Table~\ref{DNMs}. It demonstrates explicitly that
\begin{itemize}
\item[$\bullet$]one may obtain both positive (denoted by $\omega^p$) and negative (denoted by $\omega^n$) spectra for each boundary condition and, in particular, one has the following symmetry
\begin{equation}
\omega_1^n=-\omega_2^p\,,\;\;\;\;\;\;\omega_2^n=-\omega_1^p\,.\label{sym1}
\end{equation}
\item[$\bullet$]normal modes corresponding to two boundary conditions, unlike the AdS case\footnote{In an empty AdS, normal modes of Dirac fields corresponding to two boundary conditions are the same, in the sense that normal frequencies with the second boundary may be obtained by shifting $\lambda\rightarrow\lambda+1$ in the spectrum of the first boundary, except one mode~\cite{Wang:2017fie,Wang:2019qja}.}, are indeed different.
\end{itemize}
The former property indicates that, to gain the full spectra, we shall only focus on the positive spectrum with two boundary conditions; while the latter one implies that two sets of boundary conditions \textit{do} generate two distinct branches of modes.
\begin{table}
\caption{\label{DNMs} A complete normal spectra with two boundary conditions for Dirac fields in a pure cavity, with fixed $N=0$ but for different $\lambda$.}
\begin{ruledtabular}
\begin{tabular}{ l l l l l }
$\lambda$ & $\;\;\omega_1r_m$ & $\;\;\;\;\,\omega_1r_m$ & $\;\;\omega_2r_m$ & $\;\;\;\;\;\omega_2r_m$\\
\hline
1 & 3.81154 & $-$2.04279 & 2.04279 & $-$3.81154 \\
2 & 5.12311 & $-$3.20392 & 3.20392 & $-$5.12311 \\
3 & 6.37114 & $-$4.32730 & 4.32730 & $-$6.37114 \\
4 & 7.58130 & $-$5.42952 & 5.42952 & $-$7.58130 \\
5 & 8.76571 & $-$6.51789 & 6.51789 & $-$8.76571 \\
6 & 9.93123 & $-$7.59635 & 7.59635 & $-$9.93123 \\
7 & 11.0821 & $-$8.66730 & 8.66730 & $-$11.0821 \\
8 & 12.2214 & $-$9.73234 & 9.73234 & $-$12.2214 \\
9 & 13.3510 & $-$10.7926 & 10.7926 & $-$13.3510 \\
\end{tabular}
\end{ruledtabular}
\end{table}

By adding a BH in a cavity, the imaginary part of Dirac QNMs may be attained perturbatively by employing the standard matching method, which will be shown in the following.

\subsection{Analytic matching solutions}
\subsubsection{Near region solution}
In the near region and with the small charge coupling limit, by introducing a compactified coordinate $z$, $i.e.$
\begin{equation}
z\equiv \dfrac{r-r_+}{r-r_-}\;,\nonumber
\end{equation}
Eq.~\eqref{ChargedDiracU2radialR1} turns into 
\begin{equation}
z(1-z)\dfrac{d^2R_1}{dz^2}+\dfrac{1-3z}{2}\dfrac{dR_1}{dz}+\left(\hat{\omega}\dfrac{1-z}{z}-\dfrac{\bar{\lambda}^2}{1-z}\right)R_1=0 \,,\label{chargedneareq1}
\end{equation}
where
\begin{equation}
\hat{\omega}\equiv \left(\bar{\omega}+\dfrac{i}{4}\right)^2+\dfrac{1}{16}\;,\;\;\;\bar{\omega}\equiv \left(\omega-\dfrac{qQ}{r_+}\right)\dfrac{r_+^2}{r_+-r_-}\,,\nonumber
\end{equation}
and $\bar{\lambda}\equiv\lambda+\epsilon$. Here $\epsilon$ is a regulator with the order of $\omega r_+$ and, therefore, $\bar{\lambda}$ is not an exact integer anymore. Notice that although $\bar{\omega}<0$ in the superradiance regime $\omega r_+<qQ$, the final imaginary part of QNMs will not change a sign.

Eq.~\eqref{chargedneareq1} may be solved, in terms of the hypergeometric function $F(a,b,c;z)$, as
\begin{equation}
R_1\sim z^{\frac{1}{2}-i\bar{\omega}}(1-z)^{\lambda}\;F(a,b,c;z)\,,\label{chargednearsol}
\end{equation}
where an ingoing wave boundary condition at the event horizon has been imposed, and 
\begin{equation}
a=\bar{\lambda}+\frac{1}{2}\,,\;\;\;
b=\bar{\lambda}+1-2i\bar{\omega}\,,\;\;\;
c=\frac{3}{2}-2i\bar{\omega}\,.\nonumber
\end{equation}

To match with the far region solution given in the next subsection, we shall expand the near region solution, $i.e.$ Eq.~\eqref{chargednearsol}, at large $r$
\begin{align}
R_1 \;\sim \; &\dfrac{\Gamma(-2\bar{\lambda}) (r_+-r_-)^{\lambda}}{\Gamma(\frac{1}{2}-\bar{\lambda})\Gamma(1-\bar{\lambda}-2i\bar{\omega})}r^{-\lambda} \nonumber \\+&\dfrac{\Gamma(2\bar{\lambda})(r_+-r_-)^{-\lambda}}{\Gamma(\bar{\lambda}+\frac{1}{2})\Gamma(\bar{\lambda}+1-2i\bar{\omega})} r^{\lambda}
\,.\label{chargednearsolfar}
\end{align}
\subsubsection{Far region solution}
In the far region, BH effects may be neglected so the far region solution is exactly given by Eq.~\eqref{solpurecavity}. By expanding Eq.~\eqref{solpurecavity} at small $r$, we have
\begin{align}
R_1 \;\sim \; &\dfrac{\Gamma(2\lambda)}{\Gamma(1+\lambda)}\dfrac{c_2}{c_1}\left(2i\omega\right)^{-\lambda}r^{-\lambda} \nonumber \\ + &\left(1+\dfrac{\Gamma(-2\lambda)}{\Gamma(1-\lambda)}\dfrac{c_2}{c_1}\right)\left(2i\omega\right)^{\lambda} r^{\lambda}
\,,\label{chargedfarsolnear}
\end{align}
where the integration constants $c_1$ and $c_2$ may be determined by the boundary conditions. 

Supposing the cavity is placed in the far region, based on the far region solution Eq.~\eqref{solpurecavity}, one obtains
 \begin{equation}
\dfrac{c_2}{c_1}=\lambda\dfrac{\mathcal{A}_1}{\mathcal{A}_2}\,,\label{c2c1rel1}
\end{equation}
corresponding to the \textit{first} boundary condition given in Eq.~\eqref{R1bc1} and 
\begin{equation}
\dfrac{c_2}{c_1}=\lambda\dfrac{\mathcal{A}_3}{\mathcal{A}_4}\,,\label{c2c1rel2}
\end{equation}
corresponding to the \textit{second} boundary condition given in Eq.~\eqref{R1bc2}, and where
\begin{align}
\mathcal{A}_1=\;&-i M_{-\frac{1}{2}, \lambda}(2i\omega r_m)+M_{\frac{1}{2}, \lambda}(2i\omega r_m)\,,\nonumber\\
\mathcal{A}_2=\;&i\lambda W_{-\frac{1}{2}, \lambda}(2i\omega r_m)+W_{\frac{1}{2}, \lambda}(2i\omega r_m)\,,\nonumber\\
\mathcal{A}_3=\;&i M_{-\frac{1}{2}, \lambda}(2i\omega r_m)+M_{\frac{1}{2}, \lambda}(2i\omega r_m)\,,\nonumber\\
\mathcal{A}_4=\;&-i\lambda W_{-\frac{1}{2}, \lambda}(2i\omega r_m)+W_{\frac{1}{2}, \lambda}(2i\omega r_m)\,.\nonumber
\end{align}
\subsubsection{Overlap region}
In the low frequency approximation ($\omega r_+\ll1$), the near and far regions are overlapped and, by identifying Eq.~\eqref{chargednearsolfar} to Eq.~\eqref{chargedfarsolnear}, one acquires a further constraint between the integration constants $c_1$ and $c_2$, which is 
\begin{align}
&\dfrac{c_2}{c_1}=(-1)^{\lambda+1}\dfrac{\pi\bar{\omega}}{2^{2\lambda-1}}\dfrac{(\lambda-2i\bar{\omega})}{\Gamma(2\lambda)\Gamma(\lambda)}\prod^{\lambda-1}_{k=1}(k^2+4\bar{\omega}^2)\nonumber\times\\&\times\left(\coth2\pi\bar{\omega}-i\tan\pi(\lambda-\frac{1}{2})\right)\Big[\omega(r_+-r_-)\Big]^{2\lambda}\,.
\label{c2c1matching2}
\end{align}
With Eqs.~\eqref{c2c1matching2} and~\eqref{c2c1rel1} at hand, we have
\begin{align}
&\lambda\dfrac{\mathcal{A}_1}{\mathcal{A}_2}=(-1)^{\lambda+1}\dfrac{\pi\bar{\omega}}{2^{2\lambda-1}}\dfrac{(\lambda-2i\bar{\omega})}{\Gamma(2\lambda)\Gamma(\lambda)}\prod^{\lambda-1}_{k=1}(k^2+4\bar{\omega}^2)\times\nonumber\\&\times\left(\coth2\pi\bar{\omega}-i\tan\pi(\lambda-\frac{1}{2})\right)\Big[\omega(r_+-r_-)\Big]^{2\lambda}\,,\label{re1}
\end{align}
while with Eqs.~\eqref{c2c1matching2} and~\eqref{c2c1rel2} at hand, we have
\begin{align}
&\lambda\dfrac{\mathcal{A}_3}{\mathcal{A}_4}=(-1)^{\lambda+1}\dfrac{\pi\bar{\omega}}{2^{2\lambda-1}}\dfrac{(\lambda-2i\bar{\omega})}{\Gamma(2\lambda)\Gamma(\lambda)}\prod^{\lambda-1}_{k=1}(k^2+4\bar{\omega}^2)\times\nonumber\\&\times\left(\coth2\pi\bar{\omega}-i\tan\pi(\lambda-\frac{1}{2})\right)\Big[\omega(r_+-r_-)\Big]^{2\lambda}\,.\label{re2}
\end{align}

In a pure cavity, the right hand side of Eqs.~\eqref{re1} and \eqref{re2} vanish so that we have $\mathcal{A}_1=0$ and $\mathcal{A}_3=0$, which lead to normal modes exactly given in Eq.~\eqref{normalmodes}. By adding a RN BH in a cavity, one may obtain Dirac QNMs perturbatively, on top of normal modes. To do so, we assume 
\begin{equation}
\omega_j =\omega_{j,N} +i\delta_j\,,\label{modeperturb}
\end{equation}
where $\omega_{j,N}$ are the normal modes given in Eq.~\eqref{normalmodes}, and $j=1,2$ correspond to the first and second boundary conditions. Note that $\delta_j$ is generically complex and only the real part of $\delta_j$ (denoted by $\Re\delta_j$) determines the decay rate of Dirac fields and, hence, here we only focus on $\Re\delta_j$.    

By substituting Eq.~\eqref{modeperturb} into Eqs.~\eqref{re1} and~\eqref{re2}, and by expanding the corresponding results upto linear order of $\delta$ and in the limit of small $\bar{\omega}$, one obtains
\begin{equation}
\Re\delta_j=-\dfrac{\sigma_j}{4^\lambda}\dfrac{\lambda}{\Gamma(\lambda)\Gamma(2\lambda)}\prod^{\lambda-1}_{k=1}(k^2+4\bar{\omega}_{j, N}^2)\left[\omega_{j, N}(r_+-r_-)\right]^{2\lambda},\nonumber
\end{equation}
with
\[\sigma_j=\left\{\begin{array}{ll}
(-1)^{\lambda+1}\dfrac{i(1+2\lambda)\sqrt{i\omega_{1, N} r_m}}{\sqrt{2}\lambda r_m}\dfrac{\mathcal{A}_2}{\mathcal{B}_1}&\;\;\;\text{$j=1$}\,,\\\\
(-1)^{\lambda}\dfrac{i(1+2\lambda)\sqrt{i\omega_{2, N} r_m}}{\sqrt{2}\lambda r_m}\dfrac{\mathcal{A}_4}{\mathcal{B}_2}&
\;\;\;\text{$j=2$}\,,
\end{array}\right.\]
which are real and positive, and where
\begin{align}
\mathcal{B}_1=\;&(1+\lambda) M_{-1, \lambda+\frac{1}{2}}(2i\omega_{1, N} r_m)\nonumber\\&+i\lambda M_{0, \lambda+\frac{1}{2}}(2i\omega_{1, N} r_m)\,,\nonumber\\
\mathcal{B}_2=\;&(1+\lambda) M_{-1, \lambda+\frac{1}{2}}(2i\omega_{2, N} r_m)\nonumber\\&-i\lambda M_{0, \lambda+\frac{1}{2}}(2i\omega_{2, N} r_m)\,.\nonumber
\end{align}
It demonstrates explicitly that $\Re\delta_j\textless0$ is always held, no matter if the superradiance condition is satisfied, indicating Dirac fields \textit{do not} trigger superradiant instabilities. Such conclusion is held for AdS BHs as well.

These analytic calculations are only valid for the case when the cavity is placed far away from the RN BH and in the low frequency limit. In order to investigate charged Dirac QNMs in the complete parameter space, one has to resort to numerical technics, which will be addressed in the next section.

\section{Numerics}
\label{secnum}

In this section, we first briefly introduce the numeric methods employed in this paper, and then demonstrate various properties of Dirac QNMs, by taking a few selected numeric results as examples. 

\subsection{Method}
Here we present three different numerical methods, including the matrix method, the pseudospectral method and the direct integration method, to calculate Dirac QNMs on non-extremal black holes. 
\subsubsection{Matrix approach}
The matrix method is an efficient numeric approach to look for eigenvalue problems~\cite{Lin:2016sch,Lin:2017oag,Lei:2021kqv}, and here we apply this method to look for charged Dirac quasinormal spectra.

For this purpose, we first define
\begin{align}
R_1(r)=\left(r-r_+\right)^\rho\phi(r)\,,\nonumber
\end{align}
where $\rho$ is fixed as
\begin{equation}
\rho=\dfrac{1}{2}-i\dfrac{\left(\omega r_+-qQ\right)r_+}{r_+-r_-}\,,\label{bchorizon}
\end{equation}
corresponding to an ingoing wave boundary condition at the event horizon. 

Then, for numeric convenience, by introducing a dimensionless coordinate $x$
\begin{equation}
x=\dfrac{r-r_+}{r_m-r_+}\,,\nonumber
\end{equation}
which brings the integration domain from $r\in[r_+, r_m]$ to $x\in[0, 1]$, and by defining a new radial function $\chi$
\begin{equation}
\chi(x)=x \phi(x)\,,\nonumber
\end{equation}	
Eq.~\eqref{ChargedDiracU2radialR1} becomes 
\begin{equation}
\mathcal{T}_0(x,\omega)\chi(x)+\mathcal{T}_1(x,\omega)\chi'(x)+\mathcal{T}_2(x)\chi''(x)=0\,,\label{numericEq}
\end{equation}	
where $\mathcal{T}_j(j=0, 1, 2)$ may be obtained straightforwardly, and $\mathcal{T}_0$ is a quadratic function of $\omega$ while $\mathcal{T}_1$ is a linear function of $\omega$.

To solve Eq.~\eqref{numericEq}, one has to transform the boundary conditions associated to $R_1$ into the boundary conditions associated to the new radial function $\chi$. The ingoing wave boundary condition at the event horizon now turns into
\begin{equation}
\chi(0)=0\,,\label{numericbch}
\end{equation}	
while the vanishing energy flux boundary conditions at the location of a cavity, given in Eqs.~\eqref{R1bc1} and \eqref{R1bc2}, now becomes 
\begin{equation}
\dfrac{\chi'(1)}{\chi(1)}=i\left(\dfrac{K(r_m)}{\Delta(r_m)}+\dfrac{\lambda}{\sqrt{\Delta(r_m)}}\right)\Big(r_m-r_+\Big)+\rho-1\,,\label{numericbc1}
\end{equation}
and
\begin{equation}
\dfrac{\chi'(1)}{\chi(1)}=i\left(\dfrac{K(r_m)}{\Delta(r_m)}-\dfrac{\lambda}{\sqrt{\Delta(r_m)}}\right)\Big(r_m-r_+\Big)+\rho-1\,,\label{numericbc2}
\end{equation}
where $\rho$ is given in Eq.~\eqref{bchorizon}, and
\begin{equation}
\chi'(1)\equiv\dfrac{d\chi(x)}{dx}|_{x=1}.\nonumber
\end{equation}

In order to employ the matrix method, we shall first introduce equally spaced grid points in the integration domain $[0, 1]$, and then construct the differential matrices by expanding $\chi(x)$ around each grid point in the Taylor series~\cite{Lin:2016sch,Lin:2017oag}. By applying such procedures both to the radial equation~\eqref{numericEq} and to the boundary conditions~\eqref{numericbch}-\eqref{numericbc2}, they turn into algebraically matrix form
\begin{equation}
\bm{\mathcal{M}}(\omega)\bm{\chi}={\bf 0}\,,\nonumber
\end{equation}
where $\bm{\chi}$ is a vector and it is composed of the function $\chi(x)$ evaluated at each grid point, and $\bf 0$ is the zero matrix. To find out the eigenvalue $\omega$, it is nature to propose the condition 
\begin{equation}
|\bm{\mathcal{M}}(\omega)|=0\,,\nonumber
\end{equation}
where $|\bm{\mathcal{M}}(\omega)|$ is the determinant of the matrix $\bm{\mathcal{M}}(\omega)$.
\subsubsection{Pseudospectral approach}
The boundary values problems may also be alternatively solved by a pseudospectral method~\cite{trefethen2000spectral}. To do so, we shall first take the transformation
\begin{equation}
R_1(r)=\dfrac{\Delta^{1/4}}{r}e^{-i\varpi r_\ast}\chi(r)\;,\label{spectraltrans}
\end{equation}
where $r_\ast$ is the tortoise coordinate with the definition 
\begin{equation}
\dfrac{dr_\ast}{dr}=\dfrac{r^2}{\Delta}\;,\nonumber
\end{equation}
and $\varpi$ is defined as 
\begin{align}
\varpi=\dfrac{K(r_+)}{r_+^2}+\dfrac{i\Delta'(r_+)}{4r_+^2}\;.
\end{align}
Note that the transformation given in Eq.~\eqref{spectraltrans} brings an eigenvalue problem from quadratic order to linear order.

Then, for numerical purpose, we introduce the coordinate $z$ through
\begin{equation}
z=1-2\dfrac{r_m-r}{r_m-r_+}\;,\label{rtoz}
\end{equation}
so that the integration domain changes from $r\in[r_+,r_m]$ to $z\in[-1,+1]$. With the transformations~\eqref{spectraltrans} and~\eqref{rtoz} at hand, from Eq.~\eqref{ChargedDiracU2radialR1}, one obtains 
\begin{equation}
\mathcal{B}_0(z,\omega)\chi(z)+\mathcal{B}_1(z,\omega)\chi^\prime(z)+\mathcal{B}_2(z)\chi^{\prime\prime}(z)=0\;,\label{spectraleq1}
\end{equation}
where $\prime$ denotes a derivative with respect to $z$, each of the $\mathcal{B}_j (j=0,1,2)$ can be derived straightforwardly and $\mathcal{B}_j(z,\omega)=\mathcal{B}_{j,0}(z)+\omega\mathcal{B}_{j,1}(z)$ for $j=0,1$.

To solve Eq.~\eqref{spectraleq1}, one has to impose physically relevant boundary conditions for $\chi$ at asymptotic regions. At the event horizon, we impose a regular boundary condition for $\chi$ since, by considering the transformation~\eqref{spectraltrans}, an ingoing wave boundary condition is automatically satisfied for $R_1$. At the location of a cavity, on the other hand, based on Eq.~\eqref{spectraltrans}, the boundary conditions satisfied by $\chi$ are
\begin{align}
\dfrac{\chi'(1)}{\chi(1)}=\dfrac{r_m-r_+}{2}\left(\frac{1}{r_m}+i\alpha-\frac{\beta}{4\Delta(r_m)}\right)\,,\label{spectralbc}
\end{align}
where, again, $\prime$ denotes a derivative with respect to $z$, $\chi'(1)$ indicates
\begin{equation}
\chi'(1)\equiv\dfrac{d\chi(z)}{dz}|_{z=1}\;,\nonumber
\end{equation}
and
\begin{align}
&\alpha=\frac{K(r_m)}{\Delta(r_m)}+\frac{r_m^2}{r_+^2}\frac{K(r_+)}{\Delta(r_m)}\pm\dfrac{\lambda}{\sqrt{\Delta(r_m)}}\;,\label{alphaexp}\\
&\beta=\frac{r_m^2+2r_mr_+-r_+^2}{r_+}-\frac{r_m^2+r_+^2}{r_+^3}Q^2\;.\nonumber
\end{align}
Note that the $+$ and $-$ signs appeared in Eq.~\eqref{alphaexp} correspond to the first and second boundary conditions, given in Eqs.~\eqref{R1bc1} and~\eqref{R1bc2}.

By employing a pseudospectral method, one shall discretize the differential equation on a set of discrete grid points. Here we introduce the Chebyshev points and the corresponding differential matrices, which may be found for example in~\cite{trefethen2000spectral}, to discretize Eq.~\eqref{spectraleq1}. Then Eq.~\eqref{spectraleq1} turns into an algebraic equation
\begin{equation}
(M_0+\omega M_1)\chi(z)=0\;,\label{spectraleq2}
\end{equation}
where $M_0$ and $M_1$ are matrices, $(M_0)_{ij}=\mathcal{B}_{0,0}(z_i)\delta_{ij}+\mathcal{B}_{1,0}(z_i)D^{(1)}_{ij}+\mathcal{B}_{2,0}(z_i)D^{(2)}_{ij}$, and similarly for $M_1$. Note that here the first and second order Chebyshev differential matrices are denoted by $D^{(1)}$ and $D^{(2)}$. Then the quasinormal spectra may be obtained by solving Eq.~\eqref{spectraleq2}.

\subsubsection{Direct integration approach}
To solve the radial equation~\eqref{ChargedDiracU2radialR1}, we may also use the direct integration method, adapted from our previous works~\cite{Herdeiro:2011uu,Wang:2012tk,Wang:2014eha}. For this purpose, we shall first expand $R_1$ close to the event horizon
\begin{equation}
R_{1}=(r-r_+)^\rho \sum_{j=0}^\infty c_j\;(r-r_+)^j\;,\label{directexpan}
\end{equation}
where $\rho$ is given in Eq.~\eqref{bchorizon}, which indicates that an ingoing boundary condition has been imposed. Notice that here the recurrence relations among the expansion coefficients $c_j$ may be obtained directly by inserting Eq.~\eqref{directexpan} into Eq.~\eqref{ChargedDiracU2radialR1}.

Then we use Eq.~\eqref{directexpan} to initialize Eq.~\eqref{ChargedDiracU2radialR1} at the event horizon with a regulator $\epsilon$, $i.e.$ $r=r_+(1+\epsilon)$, and integrate Eq.~\eqref{ChargedDiracU2radialR1} outward from the horizon up to the location of a cavity $r_m$. By imposing the corresponding boundary conditions given in Eqs.~\eqref{R1bc1} and~\eqref{R1bc2}, one may obtain the quasinormal spectra. Note that, in practice, $\epsilon$ is normally taken from $10^{-2}$ to $10^{-5}$ until the final frequencies are stable. 

\subsection{Results}
With the above mentioned numerical methods at hand, we are ready to solve Dirac QNMs \textit{numerically}, on RN BHs in a cavity under the vanishing energy flux boundary conditions. A few selected Dirac quasinormal frequencies, with respect to the mirror radius $r_m$, the separation constant $\lambda$, the overtone number $N$, and charges of the background $Q$ and the field $q$, are presented below.

Before we proceed, a few comments are in order. In the numeric calculations, all physical quantities are measured in terms of the BH mass $M$, which amounts to set $M=1$, without loss of generality. Moreover, as we declared before, we use $\omega_1$ ($\omega_2$) to represent the quasinormal frequency corresponding to the first (second) boundary condition. Since the equation of motion is quadratic in $\omega$, each boundary condition yields two sets of modes, characterized by their real part, one being positive ($\omega^p$) and the other being negative ($\omega^n$)\footnote{Note that for charged Dirac fields on RN BHs, $\omega^p$ ($\omega^n$) denotes the frequency which originates from the spectrum with positive (negative) real part, in the neutral limit $q=0$.}. In addition, most of the numeric results presented in this part are generated by a matrix method, and they are also cross checked by the pseudospectral and direct integration approaches. We found that the results obtained by different methods are consistent. One should further notice that the illustrations on the Dirac spectrum in the following are held for both boundary conditions, unless explicitly stated otherwise. 

\subsubsection{Neutral Dirac fields on Schwarzschild BHs in a cavity}
We first explore neutral Dirac QNMs on Schwarzschild BHs in a cavity, by simply taking $Q=0$ and $q=0$.

\begin{table*}
\caption{\label{NDQNMs} A complete neutral Dirac spectrum on Schwarzschild BHs in a cavity for both boundary conditions, with fixed $N=0$ and $\lambda=1$, by varying $r_m$.}
\begin{ruledtabular}
\begin{tabular}{ l l l l l }
$r_m$ & \;\;\;\;\;\;\;\;\;\;\;\;\;\;$\omega_1$ & \;\;\;\;\;\;\;\;\;\;\;\;\;\;\;\;\;$\omega_1$ & $\;\;\;\;\;\;\;\;\;\;\;\;\;\,\omega_2$ & \;\;\;\;\;\;\;\;\;\;\;\;\;\;\;\;\,$\omega_2$\\
\hline
10 & 0.245865$-$0.030172 $i$ & $-$0.152740$-$0.004842 $i$ & 0.152740$-$0.004842 $i$ & $-$0.245865$-$0.030172 $i$ \\
9 & 0.259987$-$0.040517 $i$ & $-$0.163426$-$0.007027 $i$ & 0.163426$-$0.007027 $i$ & $-$0.259987$-$0.040517 $i$ \\
8 & 0.275231$-$0.055083 $i$ & $-$0.175312$-$0.010439 $i$ & 0.175312$-$0.010439 $i$ & $-$0.275231$-$0.055083 $i$ \\
7 & 0.291150$-$0.076020 $i$ & $-$0.188389$-$0.015888 $i$ & 0.188389$-$0.015888 $i$ & $-$0.291150$-$0.076020 $i$ \\
6 & 0.306288$-$0.106910 $i$ & $-$0.202310$-$0.024813 $i$ & 0.202310$-$0.024813 $i$ & $-$0.306288$-$0.106910 $i$ \\
5 & 0.316308$-$0.153908 $i$ & $-$0.215622$-$0.039861 $i$ & 0.215622$-$0.039861 $i$ & $-$0.316308$-$0.153908 $i$ \\
4 & 0.307692$-$0.227443 $i$ & $-$0.222965$-$0.066013 $i$ & 0.222965$-$0.066013 $i$ & $-$0.307692$-$0.227443 $i$ \\
3 & 0.231687$-$0.338217 $i$ & $-$0.202077$-$0.111119 $i$ & 0.202077$-$0.111119 $i$ & $-$0.231687$-$0.338217 $i$ \\
2.5 & 0.118629$-$0.388635 $i$ & $-$0.153429$-$0.138202 $i$ & 0.153429$-$0.138202 $i$ & $-$0.118629$-$0.388635 $i$ \\
\end{tabular}
\end{ruledtabular}
\end{table*}
By taking an example with $\lambda=1$ and $N=0$, we list a few neutral Dirac QNMs in terms of $r_m$ in Table~\ref{NDQNMs}. By observing these data, say either the first and the forth columns or the second and the third columns, one immediately uncovers the symmetry
\begin{equation}
\omega_1^n=-\left(\omega_2^p\right)^\ast\,,\;\;\;\;\;\;\omega_2^n=-\left(\omega_1^p\right)^\ast\,,\label{sym2}
\end{equation}     
which, in the pure cavity limit, turns into the symmetry given in Eq.~\eqref{sym1}. We, therefore, only pay attention to quasinormal frequencies with positive real part for two boundary conditions in the following, without loss of any information on the full spectra.  

\begin{figure}
\begin{center}
\begin{tabular}{c}
\includegraphics[clip=false,width=0.41\textwidth]{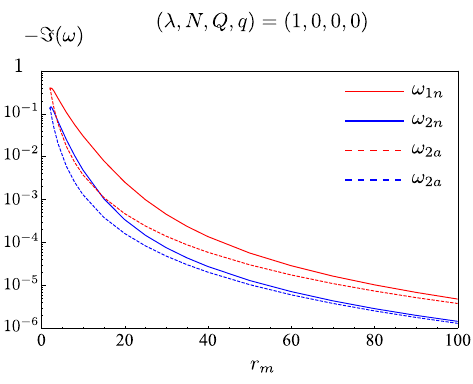}
\end{tabular}
\end{center}
\caption{\label{AVNF} (color online) A comparison for the imaginary part of Dirac QNMs between analytic (dashed lines) and numeric (solid lines) results, with the first (red) and second (blue) boundary conditions, in terms of the location of the cavity $r_m$. Notice that this figure is made with semilogarithmic coordinates.}
\end{figure}

In order to validate analytic calculations shown in section~\ref{secana} as well as to back up our numeric calculations, we present a comparison between analytic and numeric results in Fig.~\ref{AVNF}. We find, by taking $N=0$ and $\lambda=1$ as an example, a good agreement between analytic computations and numeric data, for both boundary conditions, in the regime of the analytic method is applicable, $i.e.$ when the cavity is placed far away from the hole.

\begin{figure*}
\begin{center}
\begin{tabular}{c}
\includegraphics[clip=false,width=0.42\textwidth]{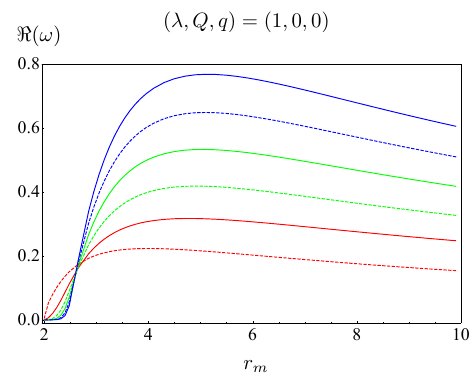}\;\;\;\;\;\;\includegraphics[clip=false,width=0.42\textwidth]{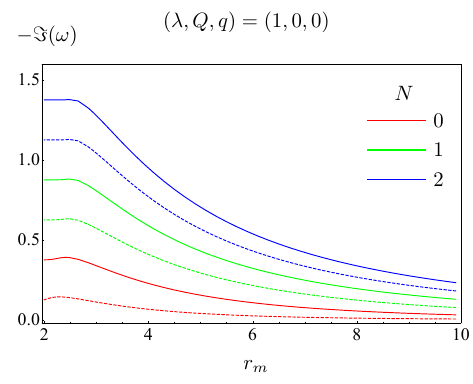}
\end{tabular}
\end{center}
\caption{\label{DiracSchradius} (color online) The neutral Dirac quasinormal spectra of the first (solid) and second (dashed) boundary conditions, with fixed $\lambda=1$ and for $N=0$ (red), $N=1$ (green) and $N=2$ (blue), in terms of mirror radius $r_m$.}
\end{figure*}

We then move to investigate the impact of $r_m$ on neutral Dirac QNMs in Fig.~\ref{DiracSchradius}. As one may observe, for all the cases we considered in this figure, the real part (the magnitude of the imaginary part) of QNMs first increases and then decreases (always increases), as $r_m$ approaches the event horizon. In particular, for all modes considered herein, when $r_m$ is close to the event horizon the quasinormal spectra become purely imaginary and approach 
\begin{equation}
\omega_1=-\left(\dfrac{N}{2}+\dfrac{3}{8}\right)i\,,\;\;\;\;\;\;\omega_2=-\left(\dfrac{N}{2}+\dfrac{1}{8}\right)i\,.\nonumber
\end{equation}    
Notice that, although the Dirac spectra become purely imaginary in this situation, the spectrum bifurcation effect does not show up, which is contradictory to the Maxwell case\footnote{For Maxwell fields on Schwarzschild BHs in a cavity, the spectrum bifurcation effect occurs when the mirror is placed close to the event horizon, characterized by quasinormal frequencies with purely imaginary parts~\cite{Lei:2021kqv}.}.

\begin{figure*}
\begin{center}
\begin{tabular}{c}
\includegraphics[clip=false,width=0.415\textwidth]{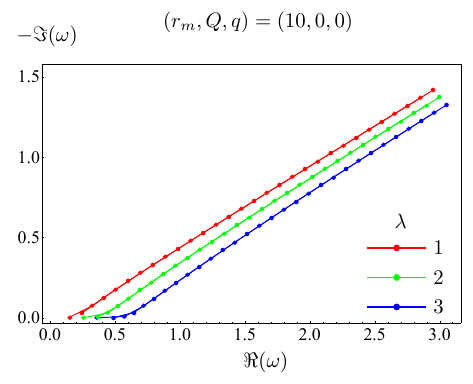}\;\;\;\;\;\;
\includegraphics[clip=false,width=0.42\textwidth]{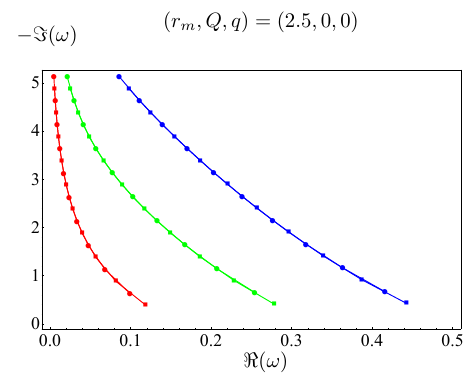}
\end{tabular}
\end{center}
\caption{\label{NDiracNSch} (color online) The neutral Dirac quasinormal spectra of the first (solid lines with circle dots) and second (solid lines with square dots) boundary conditions, with fixed $r_m=10$ (left), $r_m=2.5$ (right) and for $\lambda=1$ (red), $\lambda=2$ (green) and $\lambda=3$ (blue), in terms of the overtone number $N$.}
\end{figure*}

The impact of the overtone number $N$ on neutral Dirac QNMs, with fixed $r_m$ but for different $\lambda$, is displayed in Fig.~\ref{NDiracNSch}. Since the overtone number plays different roles in determining Dirac QNMs for different $r_m$, as already shown in Fig.~\ref{DiracSchradius}, here we take two representative mirror radius, $i.e.$ $r_m=10$ and $r_m=2.5$, to illustrate $N$ effects. For the former (latter) case, as shown in Fig.~\ref{DiracSchradius} and in the left (right) panel of Fig.~\ref{NDiracNSch}, the real part of Dirac QNMs increase (decrease) while the magnitude of imaginary part increase (increase), as $N$ increases. In particular, as shown in Fig.~\ref{NDiracNSch}, QNMs with the first and second boundary conditions lie on the same curve for different $N$, which indicates that QNMs with the second boundary may be interpolated through QNMs with the first boundary, showing some similarities for different boundary conditions in some degree.

\begin{figure*}
\begin{center}
\begin{tabular}{c}
\includegraphics[clip=false,width=0.42\textwidth]{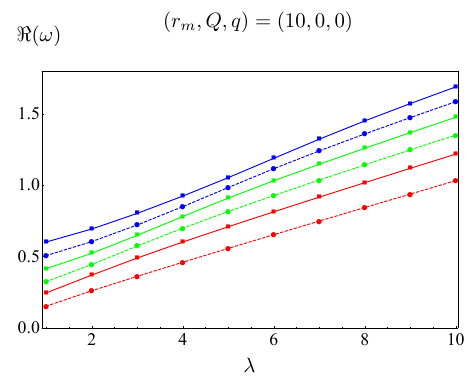}\;\;\;\;\;\;
\includegraphics[clip=false,width=0.42\textwidth]{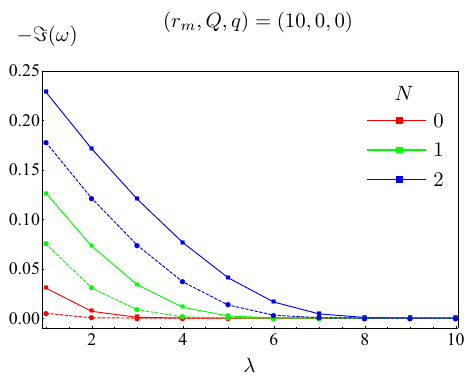}
\end{tabular}
\end{center}
\caption{\label{NDiraclambdaSch} (color online) The real (left) and imaginary (right) parts of neutral Dirac quasinormal spectra of the first (solid lines with square dots) and second (dashed lines with circle dots) boundary conditions, with fixed $r_m=10$ and for $N=0$ (red), $N=1$ (green) and $N=2$ (blue), with respect to the separation constant $\lambda$.}
\end{figure*}

The impact of the separation constant $\lambda$ on neutral Dirac QNMs, with fixed $r_m=10$ but for different $N$, is shown in Fig.~\ref{NDiraclambdaSch}. It demonstrates that, as $\lambda$ increases and for all the cases we considered herein, the real part of Dirac QNMs increases and scale (almost) linearly while the imaginary part decreases and approaches zero.  

\subsubsection{Neutral Dirac fields on RN BHs in a cavity}
In this subsection, we start to investigate the neutral Dirac QNMs, by adding charge to the background. Since the symmetry given in Eq.~\eqref{sym2} is still held for this case, here we only focus on the spectra with positive real parts for both boundary conditions. 
\begin{figure*}
\begin{center}
\begin{tabular}{c}
\includegraphics[clip=false,width=0.41\textwidth]{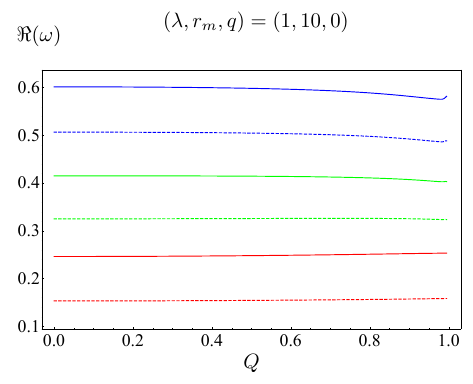}
\includegraphics[clip=false,width=0.41\textwidth]{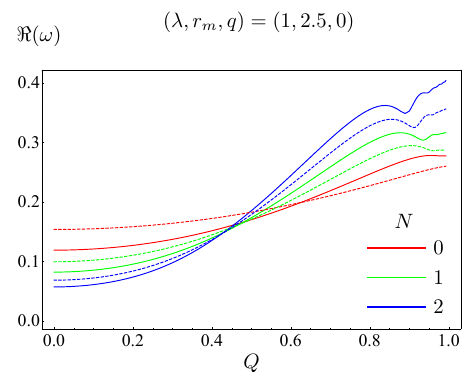}
\\
\includegraphics[clip=false,width=0.41\textwidth]{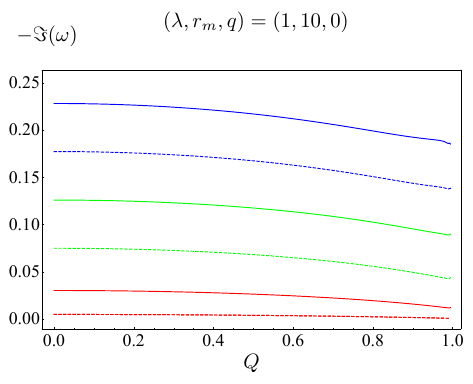}
\includegraphics[clip=false,width=0.41\textwidth]{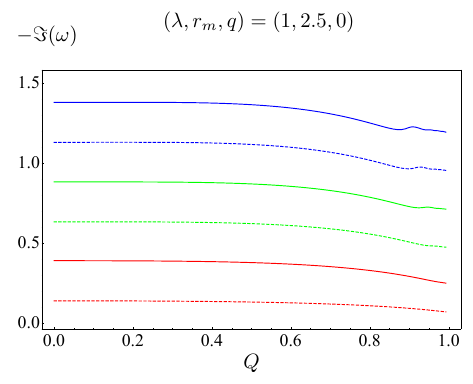}
\\
\includegraphics[clip=false,width=0.41\textwidth]{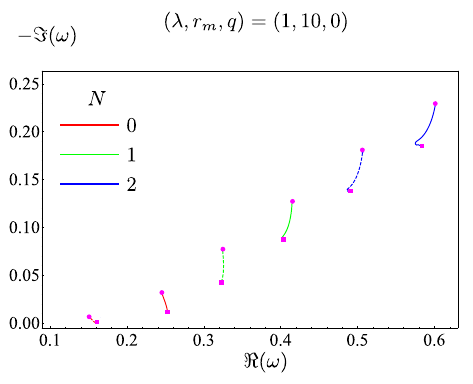}
\includegraphics[clip=false,width=0.41\textwidth]{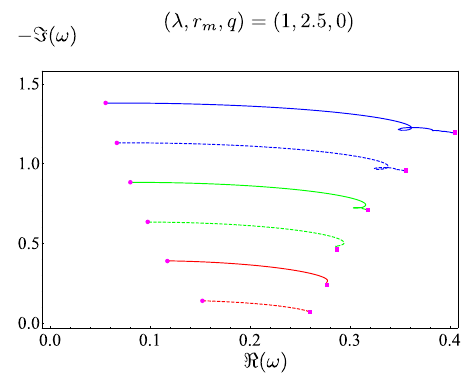}
\end{tabular}
\end{center}
\caption{\label{NDiracRNQ} (color online) QNMs of neutral Dirac fields in terms of the background charge $Q$, for the cases of $r_m=10$ (left) and $r_m=2.5$ (right), with the first (solid) and second (dashed) boundary conditions. Note that the real and imaginary parts of Dirac quasinormal spectra are shown in the first and second rows and, in the third row we display the imaginary part of QNMs with respect to the real part, by varying $Q$ from $0$ (pink dot) to $0.995$ (pink square). Here we consider the first three modes, $i.e.$ $N=0$ (red), $N=1$ (green), $N=2$ (blue), with fixed $\lambda=1$.}
\end{figure*}

To explore the impact of the background charge $Q$ on the neutral Dirac quasinormal spectra, we take two representative cases, $i.e.$ $r_m=10$ and $r_m=2.5$, with fixed $\lambda=1$ and for the first three modes, and the corresponding results are presented in Fig.~\ref{NDiracRNQ}.

For the case of $r_m=10$, as shown in the \textit{left} panel of Fig.~\ref{NDiracRNQ}, it displays that the real part of QNMs depends on $Q$ weakly. More quantitatively, as shown by the left panel in the third row, the real part of the fundamental mode $(N=0)$ increases while the excited modes $(N=1$ and $2)$ decrease, as $Q$ increases. The magnitude of the imaginary part, on the other hand, decrease for all the modes considered herein, as $Q$ increases. 

For the case of $r_m=2.5$, as shown in the \textit{right} panel of Fig.~\ref{NDiracRNQ}, it displays that, for all the modes considered herein, the real part of QNMs increase while the magnitude of the imaginary part decrease, as $Q$ increases in the most parameter space. In particular, when the background charge $Q$ is close to the charge of extremal BHs, an oscillatory pattern in the spectra may be observed for excited modes\footnote{The outburst of overtones for the case of $r_m=10$ is less obvious but true, by zooming the left panels in Fig.~\ref{NDiracRNQ} close to extremal BHs.}, which is usually dubbed as the outburst of overtones in literatures~\cite{Konoplya:2022pbc,Konoplya:2023kem,Gong:2023ghh}. This phenomenon is more pronounced for the case of $r_m=2.5$ and is expected to reflect the property and structure of the event horizon.

The impact of other parameters on the neutral Dirac QNMs for RN BHs in a cavity is quantitively similar to the counterpart for Schwarzschild BHs in a cavity shown in the last subsection and, therefore, here we do not repeat presenting those results anymore. 

\subsubsection{Charged Dirac fields on RN BHs in a cavity}
\begin{table*}
\caption{\label{CDQNMs} A complete charged Dirac quasinormal spectra on RN BHs in a cavity in terms of $q$, with fixed $N=0$, $\lambda=1$, $Q=0.8$ and $r_m=2.5$.}
\begin{ruledtabular}
\begin{tabular}{ l l l l l }
{\color{white}$-$}$q$ & \;\;\;\;\;\;\;\;\;\;\;\;\;\;\;\;\;$\omega_1$ & \;\;\;\;\;\;\;\;\;\;\;\;\;\;\;\;\;$\omega_1$ & $\;\;\;\;\;\;\;\;\;\;\;\;\;\;\;\;\omega_2$ & \;\;\;\;\;\;\;\;\;\;\;\;\;\;\;\;\,$\omega_2$\\
\hline
$-$20 & $-$6.569223$-$1.762642 $i$ & $-$7.772273$-$0.239709 $i$ & $-$6.757390$-$0.858739 $i$ & $-$8.438432$-$0.148050 $i$ \\
$-$15 & $-$4.914319$-$1.527247 $i$ & $-$5.983088$-$0.176211 $i$ & $-$5.058020$-$0.722174 $i$ & $-$6.552282$-$0.105469 $i$ \\
$-$10 & $-$3.252426$-$1.250221 $i$ & $-$4.160904$-$0.108935 $i$ & $-$3.344989$-$0.565869 $i$ & $-$4.607806$-$0.060570 $i$ \\
$-$5 & $-$1.574188$-$0.894837 $i$ & $-$2.274005$-$0.039267 $i$ & $-$1.606702$-$0.375148 $i$ & $-$2.519918$-$0.031338 $i$ \\
{\color{white}$-$}0 & {\color{white}$-$}0.246354$-$0.327664 $i$ & $-$0.226967$-$0.107303 $i$ & {\color{white}$-$}0.226967$-$0.107303 $i$ & $-$0.246354$-$0.327664 $i$ \\
{\color{white}$-$}5 & {\color{white}$-$}2.519918$-$0.031338 $i$ & {\color{white}$-$}1.606702$-$0.375148 $i$ & {\color{white}$-$}2.274005$-$0.039267 $i$ & {\color{white}$-$}1.574188$-$0.894837 $i$ \\
{\color{white}$-$}10 & {\color{white}$-$}4.607806$-$0.060570 $i$ & {\color{white}$-$}3.344989$-$0.565869 $i$ & {\color{white}$-$}4.160904$-$0.108935 $i$ & {\color{white}$-$}3.252426$-$1.250221 $i$ \\
{\color{white}$-$}15 & {\color{white}$-$}6.552282$-$0.105469 $i$ & {\color{white}$-$}5.058020$-$0.722174 $i$ & {\color{white}$-$}5.983088$-$0.176211 $i$ & {\color{white}$-$}4.914319$-$1.527247  $i$ \\
{\color{white}$-$}20 & {\color{white}$-$}8.438432$-$0.148050 $i$ & {\color{white}$-$}6.757390$-$0.858739 $i$ & {\color{white}$-$}7.772273$-$0.239709 $i$ & {\color{white}$-$}6.569223$-$1.762642 $i$ \\
\end{tabular}
\end{ruledtabular}
\end{table*}

In this subsection, we move to investigate Dirac QNMs, by adding charge to the field, on top of RN BHs.

Following the same logic as to the above subsections, we first illustrate the symmetry inherited in the spectra for charged Dirac fields. For this purpose, we list a few selected quasinormal frequencies of charged Dirac fields in Table.~\ref{CDQNMs}, by taking $N=0$, $\lambda=1$, $Q=0.8$ and $r_m=2.5$ as an example, in terms of the field charge $q$. By observing, for example, either the first and the forth columns or the second and the third columns, one may identify the symmetry for charged Dirac spectra, under the interchange of the field charge $q\rightarrow-q~$\footnote{The symmetry given in Eq.~\eqref{sym3} may be reformulated equivalently as $\omega_2^p\rightarrow-\left(\omega_1^n\right)^\ast$ and $\omega_2^n\rightarrow-\left(\omega_1^p\right)^\ast$.}, $i.e.$
\begin{equation}
\omega_1^p\rightarrow-\left(\omega_2^n\right)^\ast\,,\;\;\;\;\;\;\omega_1^n\rightarrow-\left(\omega_2^p\right)^\ast\,.\label{sym3}
\end{equation}     
In the limit of $q=0$, the symmetry satisfied by \textit{neutral} Dirac fields given in Eq.~\eqref{sym2} may be recovered. Therefore, we again only pay attention to quasinormal frequencies denoted by $\omega_1^p$ and $\omega_2^p$ in the following\footnote{Note that $\omega_1$ and $\omega_2$, appeared in all figures in this subsection, represent $\omega_1^p$ and $\omega_2^p$.}, without loss of any information on the full spectra.

\begin{figure*}
\begin{center}
\begin{tabular}{c}
\includegraphics[clip=false,width=0.42\textwidth]{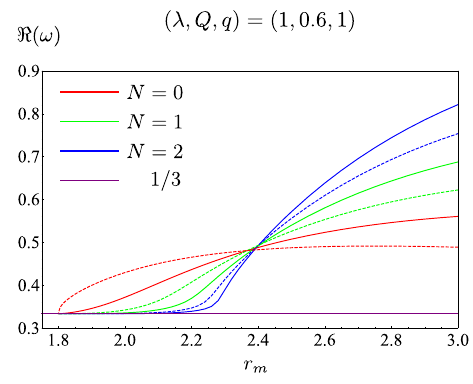}\;\;\;\;\;\;
\includegraphics[clip=false,width=0.42\textwidth]{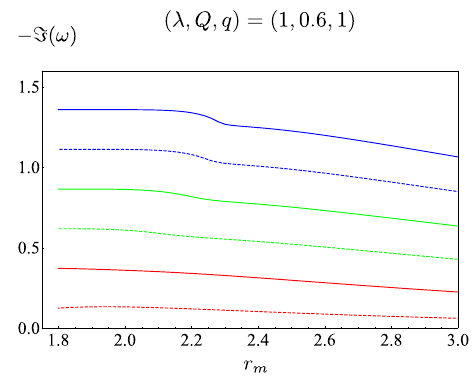}
\end{tabular}
\end{center}
\caption{\label{DiracRNradius} (color online) The real (left) and imaginary (right) parts of charged Dirac quasinormal spectra with the first (solid lines) and second (dashed lines) boundary conditions, by fixing $\lambda=1$, $Q=0.6$, $q=1$ and for $N=0$ (red), $N=1$ (green) and $N=2$ (blue), in terms of mirror radius $r_m$. Note that the horizontal line (purple) corresponds to $qQ/r_+$ which, for the case we considered herein, equals $1/3$.}
\end{figure*}
We then move to explore the Dirac quasinormal spectra in terms of the mirror radius $r_m$, by taking $\lambda=1$, $Q=0.6$, $q=1$ and for the first three modes, in Fig.~\ref{DiracRNradius}. As one may observe, the impact of $r_m$ on charged Dirac QNMs is qualitatively similar to the neutral case, as shown in Fig.~\ref{DiracSchradius}. An exceptional interesting phenomenon appeared for charged case, based on Fig.~\ref{DiracRNradius}, is that when $r_m$ approaches the event horizon $r_+$, the real part of charged Dirac QNMs approaches $qQ/r_+$, as the generalization of the neutral case.

\begin{figure*}
\begin{center}
\begin{tabular}{c}
\includegraphics[clip=false,width=0.41\textwidth]{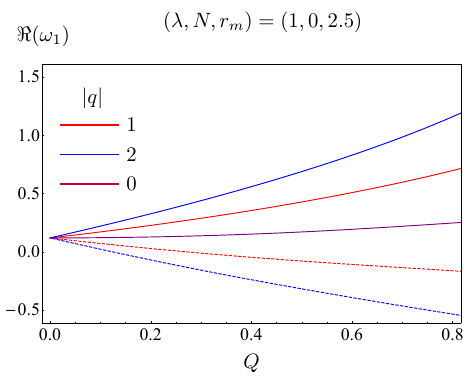}\;\;\;\;
\includegraphics[clip=false,width=0.41\textwidth]{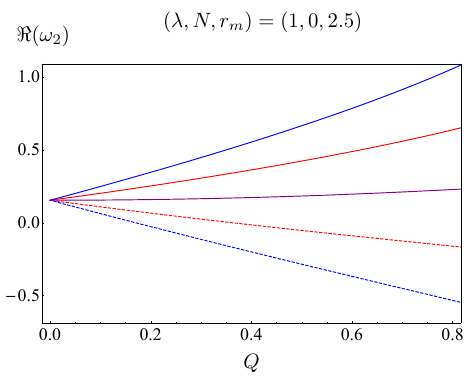}
\\
\includegraphics[clip=false,width=0.41\textwidth]{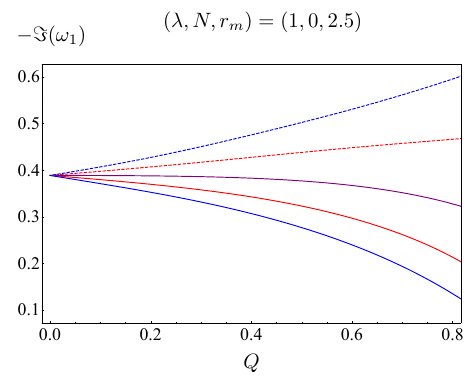}\;\;\;\;
\includegraphics[clip=false,width=0.41\textwidth]{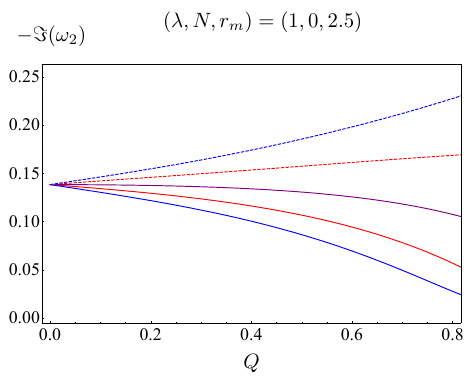}
\end{tabular}
\end{center}
\caption{\label{CDiracRNQ} (color online) Fundamental QNMs of charged Dirac fields in terms of the background charge $Q$, with fixed $r_m=2.5$ and $\lambda=1$, for the first (left) and second (right) boundary conditions. Note that here we have chosen $q=0$ (solid brown) as a reference and, $q=1$ (solid red), $q=-1$ (dashed red), $q=2$ (solid blue), $q=-2$ (dashed blue) as illustrations.}
\end{figure*}
When the field charge $q$ is present, the impact of the background charge $Q$ on the fundamental Dirac spectrum is explored in Fig.~\ref{CDiracRNQ}, with fixed $\lambda=1$ and $r_m=2.5$. As one may observe, by increasing $Q$, the real part of QNMs increases (decreases) for the case of $q\geq0$ ($q\textless0$); while the magnitude of imaginary part decreases (increases) for the case of $q\geq0$ ($q\textless0$).

\begin{figure*}
\begin{center}
\begin{tabular}{c}
\includegraphics[clip=false,width=0.41\textwidth]{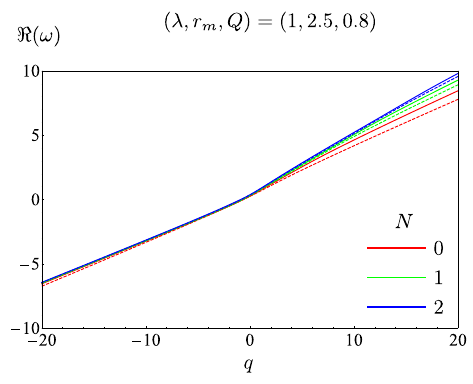}\;\;\;\;\;\;
\includegraphics[clip=false,width=0.41\textwidth]{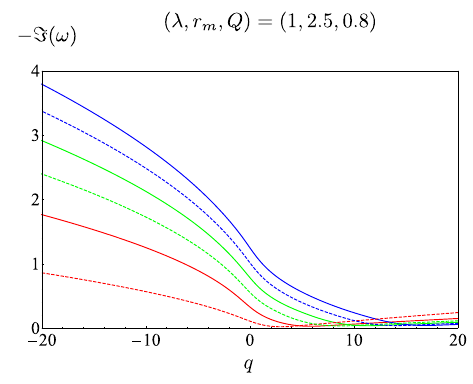}
\\
\includegraphics[clip=false,width=0.41\textwidth]{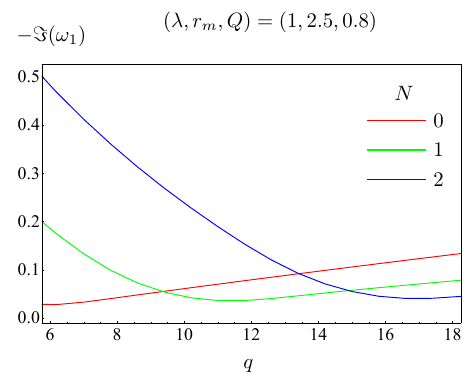}\;\;\;\;\;\;
\includegraphics[clip=false,width=0.41\textwidth]{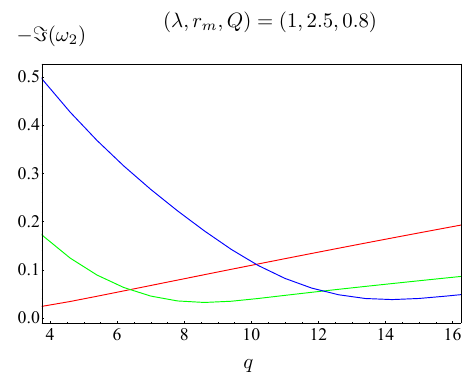}
\end{tabular}
\end{center}
\caption{\label{AnomalousDecay} (color online) QNMs of Dirac fields in terms of the field charge $q$, with fixed $\lambda=1$, $r_m=2.5$, $Q=0.8$ and for $N=0$ (red), $N=1$ (green) and $N=2$ (blue), with the first (solid) and second (dashed) boundary conditions, in the first row. In the regime of large $q$, the imaginary part of QNMs with the first (left) and second (right) boundary conditions are illustrated in the second row.}
\end{figure*}

To further explore the effects of the field charge $q$, we display charged Dirac quasinormal spectra for the first three modes in Fig.~\ref{AnomalousDecay}, by taking $\lambda=1$, $r_m=2.5$ and $Q=0.8$ as an example. As shown in the first row of Fig.~\ref{AnomalousDecay}, it demonstrates that, as $q$ increases, the real part of QNMs increases while the magnitude of the imaginary part first decreases and then increases. More than that, when $q$ is large, the anomalous decay emerges, $i.e.$ excited modes decay \textit{slower} than the fundamental mode. Such remarkable feature is more pronounced for the second boundary condition and, as far as we know, this property has \textit{not} been reported in literatures.  

\section{Discussion and Final Remarks}
\label{discussion}
In this paper we have studied charged Dirac quasinormal spectra on RN BHs in a cavity. For this purpose, we first derived equations of motion for charged Dirac fields on RN BHs, in the $\gamma$ matrices formalism. To solve the radial part of Dirac equation, one then has to impose physically relevant boundary conditions. At the event horizon, we imposed an ingoing wave boundary condition and, at the location of the cavity, we imposed two sets of explicit Robin boundary conditions obtained from the vanishing energy flux principle. With the equations of motion and boundary conditions at hand, we have explored the Dirac spectra in depth, by employing both the analytic and numeric methods.

In a pure cavity, the equation of motion for Dirac fields was solved analytically and two different sets of normal modes were obtained, generated by two sets of boundary conditions. In the limit of low frequency and small charge coupling approximations and when the cavity is placed far away from the BH, the imaginary part of charged Dirac QNMs for both boundary conditions were derived analytically, on top of normal modes, by employing the asymptotic matching approach. It was shown, similarly to the case of charged Dirac fields on RN-AdS BHs, that superradiant instabilities cannot be triggered by Dirac fields. This is the consequences of Pauli's exclusion principle at the quantum level and the violation of the area theorem at the classical level~\cite{Hawking:1974sw}. Although superradiant instability is absent for charged Dirac fields, it is still interesting to check if marginal clouds could be constructed for extremal BHs, following the analogous study of bosonic fields~\cite{Degollado:2013eqa,Sampaio:2014swa}. 

In the regime when the analytic matching method fails, we adopted the matrix method to perform numeric calculations on the Dirac quasinormal spectra. 

We have unveiled an explicit symmetry between two boundary conditions which, under the interchange of the field charge $q\rightarrow-q$, was given by $\omega_1^p\rightarrow-(\omega_2^n)^\ast$ and $\omega_1^n\rightarrow-(\omega_2^p)^\ast$, where subscripts $1$ and $2$ correspond to the first and second boundary conditions and superscripts $p$ and $n$ denote the frequency with positive and negative real parts in the neutral limit. For neutral Dirac fields, the above symmetry reduces to $\omega_1^n=-(\omega_2^p)^\ast$ and $\omega_2^n=-(\omega_1^p)^\ast$. Therefore, in order to attain the full spectra of Dirac fields, it is enough to focus on $\omega_1^p$ and $\omega_2^p$. 

By varying the mirror radius $r_m$ from a distant position to the event horizon, for both boundary conditions, it was shown that the real part of Dirac QNMs first increase and then decrease while the magnitude of the imaginary part always increase. In particular, when the cavity is placed very close to the horizon, we disclosed that, the quasinormal spectra of Dirac fields on Schwarzschild BHs become purely imaginary and asymptote to $-(3/8+N/2)i$ and $-(1/8+N/2)i$ for the first and second boundary conditions; while on RN BHs the real part of charged Dirac spectra asymptote to $qQ/r_+$ for both boundary conditions. 

By varying the overtone number $N$ and the separation constant $\lambda$, we observed the following trends. For the former case, when the cavity is placed far away from (close to) the BH and for both boundary conditions, the real part of Dirac QNMs increase (decrease) whereas the magnitude of the imaginary part increase (increase), as $N$ increases. It as especially displayed that, for both situations considered above, QNMs with the first and second boundary conditions lie on the same curve for different $N$, indicating the similarity between QNMs with two boundary conditions, in some degree. For the latter case, it was shown that, as $\lambda$ increases and for both boundary conditions, the real part of Dirac QNMs increase and scale (almost) linearly with $\lambda$ while the magnitude of the imaginary part decrease and approach zero.  

The impact of the background charge $Q$ on the fundamental Dirac QNMs is determined by the field charge $q$. We disclosed that, as $Q$ increases and for both boundary conditions, the real part of Dirac QNMs increase (decrease) for the case of $q\geq0$ ($q\textless0$) while the magnitude of the imaginary part decrease (increase) for the case of $q\geq0$ ($q\textless0$). For the case of $q=0$, we observed, among others, the existence of the outburst of overtones for excited Dirac modes. Such behaviors are more clear for the case when the cavity is placed close to the BH, providing a strong support on the argument that the outburst of overtones is sensitive to the geometry of the event horizon~\cite{Konoplya:2022pbc,Konoplya:2023kem,Gong:2023ghh}.

The impact of the field charge $q$ on Dirac spectra was also explored. It was demonstrated that,  as $q$ increases and for both boundary conditions, the real part of QNMs increase while the magnitude of the imaginary part first decrease and then increase. In particular, when $q$ is large, we observed a striking feature that excited modes decay \textit{slower} than the fundamental mode. Such anomalous decay of quasinormal spectra is expected to be the general feature for all charged fields and a further complete analysis on this phenomenon is in progress.  

By adding rotation to the background, the Dirac equations are more involved, in the sense that the radial part of Dirac equations is coupled to the angular part through the separation constant. Thus, in order to look for QNMs of Dirac fields on Kerr BHs in a cavity, one has to solve both the radial and the angular equations for Dirac fields simultaneously. Through performing such a study, it may help us not only to fully understand the Dirac spectra on BHs in a cavity but also to further verify the robustness of the vanishing energy flux boundary conditions in a much more general scenario. Work along this direction is underway and we hope to report on it in the near future.
\bigskip

\noindent{\bf{\em Acknowledgements.}}
This work is supported by the National Natural Science Foundation of China under Grant Nos. 12475050, 11705054, 12035005, by the Hunan Provincial Natural Science Foundation of China under Grant No. 2022JJ30367, by the Scientific Research Fund of Hunan Provincial Education Department Grant No. 22A0039, and by the innovative research group of Hunan Province under Grant No. 2024JJ1006. This study is also partially supported by the Innovative Experimental Projects of College Students in Hunan Normal University (2024142 and 2025304).

\bigskip

\bibliographystyle{h-physrev4}
\bibliography{DiracRNMirror} 

\end{document}